# Beyond Usability: A UX Case Study on Using "Withdrawal Design" to Challenge Engagement Metrics in Social Robotics.

YIBO MENG*, Tsinghua University, China
QIUYU LONG*, Department of Big Data Management and Application, Nanjing University of Chinese Medicine; School of Information Management, Nanjing University, China
RICHARD CHEN*, Human Centered Design & Engineering, University of Washington, United States
YAN GUAN, Arts & Design Academy, China
XIAOLAN DING†, North China University of Science and Technology Health Science Center, China

Social robots for children with autism are often evaluated through engagement and interaction quality, assuming the robot acts as a social scaffold. We report a mixed-methods "withdrawal" study that tests a harder question: what changes when the robot is removed. In an 8-week home-based randomized controlled trial (N=40), children either retained a consumer social robot (Qrobot) or had it withdrawn after initial use. Quantitatively, continued access reduced anxiety (SCARED/RCADS), yet was associated with lower parent-reported social motivation and weaker gains in emotion recognition (SMS/RMET) compared to withdrawal. Interviews with guardians contextualized this divergence: removal sometimes prompted children to seek human interaction, while continued use could keep social behavior siloed within the child–robot dyad, despite exceptionally high usability (SUS). We synthesize a UXR point of view: for vulnerable users, "engagement" can mask ecological downsides. Success should be judged not by retention, but by designed separation that bridges back to human relationships.

*The first three authors contributed equally to this research.
†Corresponding author.

Authors' Contact Information: Yibo Meng, mengyb22@mails.tsinghua.edu.cn, Tsinghua University, Beijing, China; Qiuyu Long, 012922139@njucm.edu.cn, Department of Big Data Management and Application, Nanjing University of Chinese Medicine; School of Information Management, Nanjing University, Nanjing, Jiangsu, China; Richard Chen, ruiqich@uw.edu, Human Centered Design & Engineering, University of Washington, Seattle, Washington, United States; Yan Guan, guany@tsinghua.edu.cn, Arts & Design Academy, Beijing, China; Xiaolan Ding, North China University of Science and Technology Health Science Center, Tangshan, China.

## 1 Introduction

Social robots are widely promoted as supportive tools for children with Autism Spectrum Disorder (ASD)[1–3]. In Human–Robot Interaction (HRI), they are often described as predictable, patient, and non-judgmental partners that create safer spaces for social learning[4, 5]. Within this view, the robot functions as a *social scaffold*: a temporary aid that helps children practice joint attention, turn-taking, and emotion recognition before transferring these skills to human interaction[6, 7].

This transfer assumption is central. Many studies demonstrate improved task performance while interacting with the robot itself[8, 9]. However, far fewer studies ask whether those gains persist—or generalize—beyond the robot's presence. This raises a critical question: can a scaffold become a substitute[10, 11]. If a robot provides a simplified, always-available form of social interaction, it may reduce anxiety. But it may also reduce the need to engage in more complex human relationships. The risk is not poor usability. It is ecological displacement. A system designed to support connection could unintentionally narrow it[12–14].

To examine this tension, we conducted an 8-week home-based randomized controlled trial of a consumer social robot (Qrobot). Our key methodological move was simple but uncommon: we included a withdrawal condition. After initial use, one group retained the robot, while the other had it removed. Rather than measuring performance only in the robot's presence, we examined what changed when it was gone.

We hypothesized that continued robot use would improve social motivation (H1) and emotion recognition/empathy (H2), and reduce anxiety (H3). Our findings complicate this expectation. The robot significantly reduced anxiety. Yet continued access was associated with lower social motivation and weaker gains in emotion recognition compared to withdrawal. Despite extremely high usability scores, the robot did not consistently function as a bridge to human interaction.

We argue that HRI needs to look beyond engagement and interaction quality. For vulnerable users, success may not be defined by retention, but by whether the system helps users re-engage with the human world.

*Contributions.* This work contributes a strategic UXR play, the *Withdrawal Test*, for evaluating habit-forming technologies in sensitive contexts. By comparing retention and removal conditions in a natural home environment, we show how high usability and reduced anxiety can mask ecological trade-offs in social motivation and emotional development. We further propose a shift in product evaluation for vulnerable users, from maximizing engagement to designing for transfer and eventual separation.

## 2 Related Works

### 2.1 Robots as Social Scaffolds

In ASD interventions, social robots are commonly framed as scaffolds[10, 11]. The idea is straightforward: robots simplify social interaction. They are structured, predictable, and easier to interpret than humans[15, 16]. This reduced complexity is believed to lower cognitive load and create a supportive learning environment.

Empirical studies show that robots such as Kaspar, NAO, and Milo can elicit joint attention, imitation, verbal initiation, and turn-taking in controlled settings[17, 18]. These findings support the scaffold model: skills practiced with the robot may transfer to human–human interaction[19].

However, most evaluations focus on performance during robot interaction. Fewer studies examine long-term generalization in natural home environments[1, 20–22]. Whether scaffolding reliably leads to transfer remains underexplored.

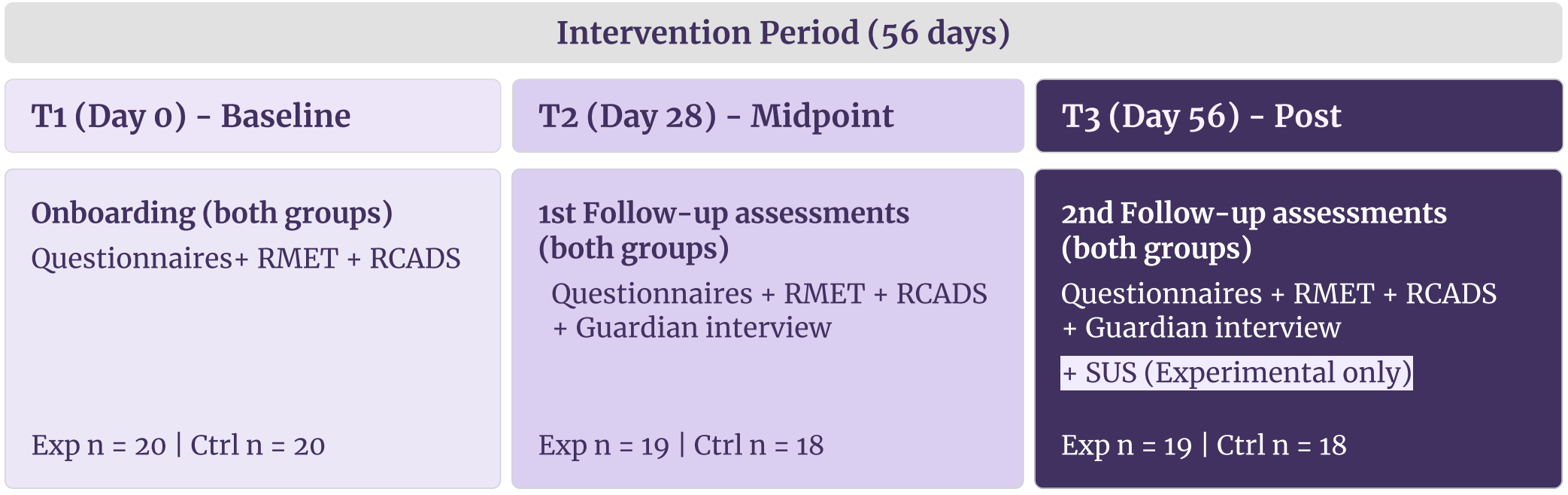

Fig. 1. Study timeline illustrating the withdrawal design across T1, T2, and T3.

## 2.2 Robots as Anxiolytic Agents

Prior HRI studies consistently report reduced anxiety or physiological arousal in children with ASD during robot interaction[23]. Given that children with ASD often experience stress in unpredictable social contexts[24], a structured and rule-based partner may enhance predictability and perceived control, supporting short-term emotional regulation.

Both physiological and questionnaire-based measures report decreased anxiety during or after robot interaction[24]. This effect is often interpreted as beneficial. Lower anxiety may enable participation and learning[25]. Yet this raises a tension. If the robot becomes the most comfortable social partner, does reduced anxiety encourage human engagement—or replace it?

## 2.3 From Scaffold to Substitute

Recent critical perspectives in HRI urge caution regarding long-term ecological effects[26, 27]. Many studies are shortterm and laboratory-based[28, 29]. Less is known about how robots reshape daily relational dynamics in homes. This concern has led to what we refer to as the substitution hypothesis[30, 31]. A robot designed as a bridge may instead become a preferred endpoint. If it provides predictable, low-demand interaction, it may reduce the motivation to engage in more effortful human relationships[32, 33]. This aligns with broader critiques about technological companionship that simulates connection without reciprocity[34]. Empirically testing substitution requires more than observing interaction quality. It requires examining what happens when the robot is removed. To our knowledge, withdrawal designs are rare in HRI studies of ASD. Our work addresses this methodological gap.

# 3 Methods

To test the scaffolding-to-substitution hypothesis, we conducted an 8-week home-based randomized controlled trial (RCT) with a withdrawal condition. Our goal was to compare outcomes under continued robot access versus removal, and to examine not only what happens during human–robot interaction, but what changes when the robot is absent. We used a mixed-methods design combining validated psychometric measures with semi-structured guardian interviews.

## 3.1 Participants

Families were recruited through online channels (Weibo, RedNote, WeChat) and community outreach. Children were eligible if they (1) were aged 5–10, (2) had a formal clinical diagnosis of ASD, and (3) had prior experience with Qrobot for no more than three months. We set this usage window to ensure children were familiar with the robot while reducing ethical risk associated with long-term dependency during withdrawal. Additional inclusion criteria included basic receptive language ability, child assent when appropriate, and written guardian consent.

Children were excluded if they had severe sensory or neurological comorbidities, significant aggressive or selfinjurious behaviors that would prevent safe participation, concurrent enrollment in other social-intervention studies within the past six months, or experience with other social robot systems.

### 3.2 Study Design and Procedure

Figure 1 illustrates the temporal structure of the withdrawal design across the 8-week intervention.

The study included three assessment points: baseline (T1), midpoint at week 4 (T2), and post-intervention at week 8 (T3). **Baseline (T1).** After onboarding and consent, guardians completed the Social Motivation Scale (SMS)[35], Basic Empathy Scale (BES)[36], and Screen for Child Anxiety Related Emotional Disorders (SCARED)[37]. Children completed the Revised Children's Anxiety and Depression Scale (RCADS)[38] and the Reading the Mind in the Eyes Test, Child Version (RMET)[39].

**Randomization and intervention.** After T1, participants were randomly assigned to one of two arms:

- Retain group: families continued their usual Qrobot use.
- Withdrawal group: Qrobot was removed from the home after baseline.

**Midpoint (T2) and post-intervention (T3).** At T2 and T3, guardians repeated the SMS, BES, and SCARED, and children repeated the RCADS and RMET. Guardians also participated in semi-structured interviews at both time points. At T3, children in the retain group completed the System Usability Scale (SUS)[40] to assess perceived ease of use and to contextualize outcomes relative to interaction quality.

### 3.3 Measures

We selected measures to capture three domains central to the scaffold-versus-substitute question: social motivation, emotion understanding, and anxiety.

**Social motivation.** Guardians completed the SMS[35] to assess the child's social approach, maintenance, and enjoyment.

**Emotion recognition and empathy.** Children completed the RMET (Child Version)[39]. Guardians completed the BES[36], covering cognitive and affective empathy.

**Anxiety.** Guardians completed SCARED[37]. Children completed RCADS[38]. Using both informant- and self-report measures allowed us to capture anxiety-related change from complementary perspectives.

**Usability.** Children in the retain group completed SUS at T3[41]. We used SUS to verify that any observed downstream effects were unlikely to be driven by poor usability.

### 3.4 Qualitative Interviews

We conducted semi-structured interviews with each child's primary guardian at T2 and T3. Interviews elicited concrete examples of changes in: (1) social initiation and engagement with family or peers; (2) emotion understanding and empathic behaviors in daily life; (3) emotional stability and anxiety-related behaviors; and (4) how Qrobot fit into household routines. Interview prompts emphasized observable events (e.g., "Can you describe a recent situation where your child tried to join another person?") rather than global judgments.

All interviews were conducted online (Tencent Meeting or Zoom), audio-recorded with consent, and transcribed verbatim. Transcripts were checked for accuracy by at least two researchers.

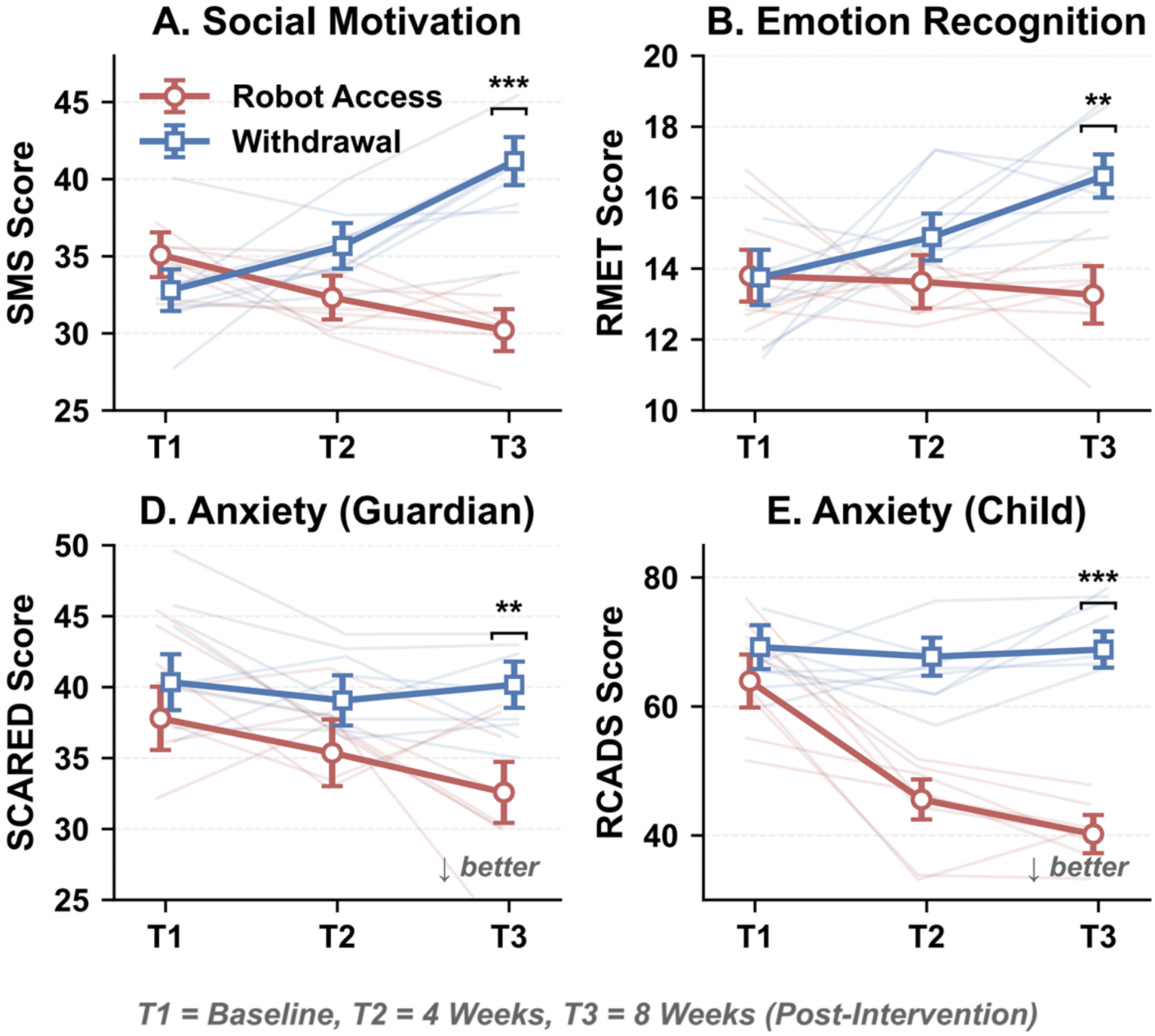

Fig. 2. Divergent outcome trajectories under continued robot access versus withdrawal. (A) Social motivation (SMS) increased in the withdrawal group but declined under continued robot access. (B) Emotion recognition (RMET) improved following withdrawal but remained flat or decreased with retention. (D–E) Anxiety (guardian-reported SCARED; child-reported RCADS) decreased under continued robot access, with minimal change following withdrawal. Points represent group means (± SE); faint lines indicate individual trajectories. T1 = Baseline, T2 = 4 weeks, T3 = 8 weeks (post-intervention). Asterisks denote between-group differences at T3.

### 3.5 Analysis

**Quantitative.** We summarized each measure using means and standard deviations at T1, T2, and T3. Group differences were evaluated using two-tailed tests with $\alpha = .05$. Our primary comparisons focused on between-group differences at T3 and on change over time.

**Qualitative.** We analyzed interviews using thematic analysis. Researchers iteratively coded transcripts, compared interpretations, and refined themes around (1) social transfer versus social siloing, (2) emotion understanding in real contexts, and (3) anxiety regulation. We then integrated qualitative themes with quantitative patterns to explain not only *whether* outcomes differed across groups, but *how* families experienced these changes in everyday life.

## 4 Results

Forty-four participants were enrolled; forty completed the full 8-week study (retain group: $n = 19$; withdrawal group: $n = 21$). The mean age was 6.78 years ($SD = 1.77$). Demographic characteristics are summarized in Appendix C. Figure 2 presents longitudinal trajectories for social motivation (SMS), emotion recognition (RMET), and anxiety (SCARED, RCADS) across baseline (T1), week 4 (T2), and week 8 (T3). We integrate quantitative and qualitative findings below to illustrate how behavioral trajectories diverged between continued robot access and withdrawal.

### 4.1 Insight 1: Reduced Anxiety Under Continued Robot Use

Continued access to Qrobot was associated with significant reductions in anxiety.

On the guardian-reported SCARED, the retain group showed lower anxiety scores at T3 compared to the withdrawal group ($t = -2.762, p = .009$). A similar pattern emerged in child self-reports on the RCADS. The retain group showed a marked decrease from T1 ($M = 63.95$) to T3 ($M = 40.21$), while the withdrawal group remained relatively stable ($t = -7.003, p < .001$). Longitudinal trajectories are shown in Figure 2.

Interview data aligned with these patterns. Guardians in the retain group frequently described shorter tantrums, smoother emotional recovery, and calmer responses during structured activities. One mother (P6) noted:

> *Before, if something upset him, he might cry and scream for over half an hour. Now he can calm down in about twenty minutes. It's not dramatic, but I can feel the difference.*

Parents also described the robot as creating a positive emotional tone that extended beyond the interaction itself (P11).

In contrast, guardians in the withdrawal group reported no sustained change in anxiety after an initial brief adjustment period. Several parents described short-lived disappointment when the robot was removed, followed by emotional stabilization (P29, P39).

### 4.2 Insight 2: Lower Social Motivation With Continued Access

Despite reduced anxiety, continued robot access was associated with lower parent-reported social motivation at T3.

On the Social Motivation Scale (SMS), both groups began at similar baselines. By T3, the retain group scored significantly lower ($M = 30.21, SD = 5.92$) than the withdrawal group ($M = 37.22, SD = 4.67; t = -3.984, p < .001$).

Qualitative accounts reflected this pattern. Guardians in the withdrawal group frequently described increases in spontaneous social initiation. One mother (P25) explained:

> *After the first two weeks, he started to seek more interaction with us. He would bring a book and say, "Mommy, read." That was rare before.*

Another parent (P39) described a shift toward more direct verbal requests instead of gesture-based communication.

In contrast, guardians in the retain group often described interaction remaining concentrated around the robot. One mother (P4) observed:

> *When guests visit, he retreats to play with the robot. He doesn't greet them or show them what he learned.*

Another parent (P9) reported that attempts to use the robot as a bridge to peer interaction did not persist once the peer left.

### 4.3 Insight 3: Divergent Patterns in Emotion Recognition

Emotion recognition outcomes also differed between groups.

On the RMET, the withdrawal group improved by T3 ($M = 16.61, SD = 2.57$), while the retain group slightly declined ($M = 13.26, SD = 3.54$), producing a significant between-group difference ($t = -3.276, p = .002$; Figure ??). Patterns on the BES followed a similar directional trend.

Interview data suggested differences in how emotion understanding appeared in daily life. Guardians in the withdrawal group described children responding more sensitively to tone and non-verbal cues. For example, one mother (P28) recounted:

> *She saw I had a headache, brought a blanket, and said, "Mommy, rest." That was new.*

Another parent (P27) described improved recognition of emotional tone in voice.

In the retain group, guardians often described emotion knowledge that remained tied to robot-led exercises. One father (P12) noted:

> *He can answer the robot's questions about emotions in pictures, but when his sister cries, he seems unsure what to do.*

Several parents described repeated emotion labeling without clear contextual application (P13).

### 4.4 System Usability

At T3, the retain group reported exceptionally high usability (SUS: $M = 98.16, SD = 4.20$). Participants consistently rated Qrobot as easy to use and straightforward to interact with.

Taken together, the findings show a clear pattern: continued robot use was associated with reduced anxiety but lower social motivation and weaker gains in emotion recognition compared to withdrawal.

## 5 Discussion

This study set out to test a simple but underexplored question: does a social robot function as a bridge to human interaction, or can it become a destination in itself?

Our findings show a consistent tension. Continued access to the robot reduced anxiety. At the same time, children in the withdrawal group showed stronger gains in social motivation and emotion recognition. The same system that stabilized emotions did not consistently support social transfer.

### 5.1 From Scaffold to Substitute

The most consequential finding concerns social motivation. Children who retained the robot showed lower SMS scores at T3 compared to those in the withdrawal group. Interview data suggest that, for some children, interaction remained concentrated within the child–robot dyad.

In contrast, removal of the robot was followed by increased spontaneous social initiation in several families. Parents described a shift from object-centered interaction to person-directed interaction. This pattern is consistent with research suggesting that environmental constraints can reshape behavior and attention[42–44].

We interpret this not as evidence that robots are harmful, but as evidence that scaffolding does not automatically produce transfer. Without structured mechanisms that redirect engagement toward humans, a predictable interaction partner may reduce the incentive to navigate more demanding social contexts.

### 5.2 Emotion Knowledge Without Generalization

A similar divergence appeared in emotion recognition. The withdrawal group improved on the RMET, while the retain group did not. Guardian reports suggest that emotion knowledge in the retain group often remained tied to scripted robot exercises.

Decoding real human emotion requires variability, ambiguity, and reciprocal feedback[45–48]. Simplified emotional cues may support labeling, but may not foster flexible interpretation across contexts. The issue appears to be not whether children can learn emotion terms, but whether that learning generalizes beyond the training environment.

### 5.3 The Paradox of Predictability

The robot's anxiolytic effect was robust. Both SCARED and RCADS showed significant reductions in the retain group. Interview data consistently described shorter tantrums and smoother emotional recovery.

This aligns with prior work highlighting the regulatory value of predictable, structured robot interaction[49, 50]. Predictability can reduce uncertainty and cognitive load.

However, predictability may also limit exposure to the variability inherent in human social life. The same features that create emotional safety may also narrow social exploration. This is the central tension revealed by our data.

### 5.4 Implications for HRI Design

Taken together, these findings suggest a reframing of success metrics in HRI for vulnerable populations[51]. High usability and high engagement are not sufficient indicators of positive long-term impact. In our study, SUS scores were exceptionally high, yet social outcomes diverged.

We propose a shift in design orientation:

(1) **Design for transfer, not retention.** Systems should include explicit mechanisms that guide interaction outward—toward caregivers or peers—rather than reinforcing exclusive dyadic engagement[23, 52]. Gradual fading, shared tasks, or co-use scenarios may help reposition the robot as a bridge.
(2) **Reposition caregivers as active participants.** Robots should scaffold caregiver–child interaction, not replace it[53–56]. Embedding prompts, joint activities, or structured handoffs may increase generalization and sustainability.

More broadly, our findings support an ecological view of HRI[57, 58]. Robots do not operate in isolation. They reshape attention, routines, and relational dynamics within the home.

### 5.5 Limitations and Future Directions

This study has limitations. The sample size was modest and regionally recruited. The intervention lasted eight weeks, which does not capture long-term adaptation. Our measures relied primarily on questionnaires and interviews rather than fine-grained behavioral logs.

Future work should examine differentiated designs tailored to age and symptom profile[59–62], integrate multimodal behavioral measures[63–68], and conduct longer-term studies that include planned fading phases[69, 70]. Understanding how robots transition from introduction to withdrawal may be as important as understanding their initial effects.

## 6 Conclusion

This study demonstrates a dual effect of social robots in home-based ASD contexts. Continued use reduced anxiety but did not consistently support social transfer. Withdrawal, in contrast, was associated with increased social initiation and improved emotion recognition in several cases.

We argue that the key design question is not whether robots engage children, but whether they help children re-engage with people. For vulnerable users, success may be defined not by sustained attachment to the system, but by the system's ability to become unnecessary.

## Generative AI Use Disclosure

During the preparation of this manuscript, we utilized large language models in a limited capacity solely for editorial refinement and consistency checks. All authors assume full responsibility for the content, analysis, and interpretations.

## A Appendix

*Table S1.* System Usability Scale (SUS) Questionnaire.

Table 1. System Usability Scale (SUS) Questionnaire.

| Question Number | Statement | 1-5 Points |
|---|---|---|
| 1 | I think that I would like to use this system frequently. | |
| 2 | I found the system unnecessarily complex. (Reverse score) | |
| 3 | I thought the system was easy to use. | |
| 4 | I think that I would need the support of a technical person to use this system. (Reverse score) | |
| 5 | I found the various functions in this system were well integrated. | |
| 6 | I thought there was too much inconsistency in this system. (Reverse score) | |
| 7 | I would imagine that most people would learn to use this system very quickly. | |
| 8 | I found the system very cumbersome to use. (Reverse score) | |
| 9 | I felt very confident using the system. | |
| 10 | I needed to learn a lot of things before I could get going with this system. (Reverse score) | |

## B Appendix

*Table S2.* Semi-Structured Interview Protocol.

## C Appendix

*Table S3.* Participant Demographic Information.

Table 2. Semi-Structured Interview Protocol.

| Interview Module | Core Objective | Questions & Guides |
|---|---|---|
| Basic Information | User Profile | Age, Gender,<br>Urban/Rural, ASD History,<br>Relationship to the Child |
| Overall Experience & Acceptability | To understand the degree of Qrobot's integration into daily life and basic user experience. | 1. Please describe when and in what situations your child typically used Qrobot over the past 8 weeks.<br>2. Overall, what was your child's attitude towards Qrobot? (e.g., Did they look forward to it? Like it?)<br>3. Did you find the Qrobot itself easy to operate? What difficulties, if any, did you encounter while guiding your child to use it? |
| Impact on Social Motivation & Behavior | To explore specific behavioral changes in the child's social initiation, maintenance, and enjoyment. | 4. (Social Initiation) Have you observed any changes in your child's willingness to initiate social interactions after using Qrobot? For example, initiating conversations, greeting family members, or asking questions? Please provide an example.<br>5. (Social Maintenance) Have you noticed any changes in your child's ability to sustain a conversation or interaction? For example, can they engage in longer exchanges or take turns in games?<br>6. (Social Enjoyment) Do you feel there has been a change in the pleasure your child derives from interacting with you or others? Do they seem to enjoy socializing more? |
| Impact on Emotion & Empathy | To assess subtle changes in the child's ability to recognize, express emotions, and show empathetic responses. | 7. (Emotion Recognition) Qrobot includes content about emotions. Do you feel this has helped your child in recognizing their own or others' emotions (like happiness or sadness)? Please explain.<br>8. (Empathetic Behavior) Have you noticed your child displaying more empathetic behaviors? For instance, showing concern or offering comfort when they see someone is upset? |
| Impact on Psychological State & Anxiety | To understand changes in the child's emotional stability and anxiety levels in specific situations. | 9. Overall, how would you describe your child's emotional stability during these 8 weeks? Compared to before, have there been any changes in their anxiety, nervousness, or frequency of tantrums?<br>10. In which specific situations (e.g., facing a new environment, task challenges) did you find Qrobot's help most noticeable? Or least noticeable? |
| Overall Evaluation & Suggestions | To gather overall guardian evaluations and improvement suggestions for optimizing the intervention program. | 11. Setting the research requirements aside, what do you personally feel was the most significant help Qrobot provided for your child? (This can be in any aspect).<br>12. What shortcomings do you think Qrobot has, or in what areas could it be improved?<br>13. If possible, would you be willing to let your child continue using Qrobot? Why or why not? |

Table 3. Participant Demographic Information.

| Participant ID | Age | ASD History (Months) | Gender | Residence | Guardian in Study | Group |
|---|---|---|---|---|---|---|
| 1 | 7 | 11 | F | U | Mother | Experimental |
| 2 | 7 | 12 | F | R | Father | Experimental |
| 3 | 7 | 22 | F | U | Mother | Experimental |
| 4 | 6 | 12 | M | R | Mother | Experimental |
| 5 | 9 | 13 | M | R | Father | Experimental |
| 6 | 5 | 14 | M | R | Mother | Experimental |
| 7 | 9 | 21 | F | R | Father | Experimental |
| 8 | 5 | 22 | F | R | Mother | Experimental |
| 9 | 7 | 15 | M | R | Father | Experimental |
| 10 | 7 | 14 | F | U | Mother | Experimental |
| 11 | 7 | 2 | M | R | Father | Experimental |
| 12 | 9 | 7 | M | U | Father | Experimental |
| 13 | 8 | 17 | M | R | Mother | Experimental |
| 14 | 7 | 13 | M | R | Mother | Experimental |
| 15 | 7 | 9 | F | U | Grandmother | Experimental |
| 16 | 6 | 9 | M | U | Mother | Experimental |
| 17 | 6 | 11 | F | R | Mother | Experimental |
| 18 | 7 | 12 | F | R | Mother | Experimental |
| 19 | 6 | 33 | M | R | Mother | Experimental |
| 20 | 7 | 34 | F | U | Mother | Experimental |
| 21 | 5 | 21 | M | R | Mother | Control |
| 22 | 5 | 12 | F | U | Mother | Control |
| 23 | 9 | 14 | F | R | Father | Control |
| 24 | 5 | 17 | M | U | Grandmother | Control |
| 25 | 5 | 15 | M | R | Mother | Control |
| 26 | 5 | 14 | M | U | Mother | Control |
| 27 | 5 | 21 | M | R | Mother | Control |
| 28 | 9 | 22 | F | R | Mother | Control |
| 29 | 5 | 15 | F | R | Mother | Control |
| 30 | 7 | 14 | F | U | Mother | Control |
| 31 | 7 | 19 | M | U | Mother | Control |
| 32 | 7 | 22 | F | U | Father | Control |
| 33 | 7 | 24 | M | R | Mother | Control |
| 34 | 7 | 16 | F | R | Father | Control |
| 35 | 6 | 14 | M | U | Mother | Control |
| 36 | 7 | 21 | F | R | Father | Control |
| 37 | 9 | 22 | F | U | Father | Control |
| 38 | 9 | 34 | M | R | Mother | Control |
| 39 | 6 | 34 | F | U | Father | Control |
| 40 | 7 | 45 | M | R | Grandmother | Control |